\documentstyle[multicol,aps,epsfig]{revtex}
\begin{document}

\title{High-field limit of the
Vlasov-Poisson-Fokker-Planck system: A comparison
of different perturbation methods}
\author{Luis L. Bonilla}
\address{Escuela Polit\'ecnica Superior,
Universidad Carlos III de Madrid, Avda.\
Universidad 30, \\ 28911 Legan{\'e}s, Spain.
E-mail: {\tt bonilla@ing.uc3m.es}}
\author{Juan Soler}
\address{Departamento de Matem{\'a}tica Aplicada, Facultad de Ciencias,
Universidad de Granada \\18071 Granada, Spain.
E-mail: {\tt jsoler@ugr.es}}

\date{June 30, 2000}

\maketitle
\begin{abstract}
A reduced drift-diffusion (Smoluchowski-Poisson)
equation is found for the electric charge in the
high-field limit of the
Vlasov-Poisson-Fokker-Planck system, both in one
and three dimensions. The corresponding electric
field satisfies a Burgers equation.
Three methods are compared in the one-dimensional
case: Hilbert expansion, Chapman-Enskog procedure
and closure of the hierarchy of equations for the
moments of the probability density. Of these
methods, only the Chapman-Enskog method is able to
systematically yield reduced equations containing
terms of different order.
\end{abstract}


\begin{multicols}{2}
\narrowtext

\section{Introduction}
The Vlasov-Poisson-Fokker-Planck (VPFP) system
describes the probability density $\rho$ of an ensemble
of electrically charged Brownian particles in
contact with a thermal bath and subject to their
self-consistent electric field $\phi$, see \cite{bon97}. It
consists of the equations:
\begin{eqnarray}
\frac{\partial \rho}{\partial t}
+ v\cdot\nabla_{x}\rho - \nabla_{v}\cdot
\left[\left(\mu v + {e\over m}\nabla_x \phi
\right)\,\rho \right. \nonumber\\
\left. +
\frac{\mu k_{B}T}{m}\,\nabla_v \rho \right] =
0 \,,   \label{vpfpe}\\
\varepsilon\, \Delta_x \phi =
- e N_0 \int \rho\, dv, \label{poisson}\\
\int \rho(x,v,t)\, dx\, dv = 1. \label{norma}
\end{eqnarray}
Here $N_0$ is the number of particles, $\mu$ a
friction coefficient, $e$ and $m$ particle charge
and mass, $T$ temperature and $\varepsilon$
permittivity. Sometimes it is interesting to
consider the high-field limit, in which the
collision term (divergence with respect to $v$)
in (\ref{vpfpe}) dominates the other two. In this
case, $\rho$ quickly evolves toward a displaced
Maxwellian, 
$$
\left( {m\over 2\pi k_{B}T}\right)^{{3\over 2}}
\exp\left[ - {m
\left(v+ {e\over m\mu}
\nabla_{x}\phi\right)^{2}\over 2k_{B}T}\right] 
$$
times a slowly-varying function, $P(x,t)$.
Essentially, $P$ is a charge density whose
evolution we would like to describe. Appropriate
scales in which this fast equilibration to the
displaced Maxwellian occurs may be found as
follows. A typical velocity is found by balancing
friction, $\mu v \rho$ and flux
$\mu k_B T \nabla_v \rho/m$, which yields $\mu
v_0 = \mu k_B T/(mv_0)$, or
\begin{equation}
v_0 = \sqrt{{ k_{B}T\over m}}\,,\label{v0}
\end{equation}
which is the thermal velocity. If we impose that
$\mu v$ and $e\nabla_x \phi/m$ be of the same
order, we find the unit of electric field, $E_0$,
\begin{equation}
E_0 = {\mu m v_{0} \over e} = {\mu v_{0}\over
e}\, \sqrt{m k_{B}T}\,.\label{E0}
\end{equation}
The Poisson equation yields the unit of length,
$\varepsilon E_0/ x_0 = e N_0/x_0^3$, or
\begin{equation}
x_0 =  \sqrt{{eN_{0}\over\varepsilon E_{0}}} = e
\left({N_{0}\over\mu\varepsilon}\right)^{{1\over
2}} \, (mK_B T)^{-{1\over 4}}\,.
\label{x0}
\end{equation}
The relation between the second and third terms
in (\ref{vpfpe}) yields the small parameter
\begin{equation}
\epsilon = { v_{0} \over \mu x_{0}} = {1\over
e}\, \left({\varepsilon^{2} (k_{B}T)^{3}\over m
(\mu N_{0})^{2} } \right)^{{1\over 4}}\,.
\label{epsilon}
\end{equation}
The unit of $\rho$ should be $(x_0 v_0)^{-3}$
because of the normalization condition
(\ref{norma}). Finally, the unit of time should
be $x_0/v_0$ provided we want the first two terms
in (\ref{vpfpe}) to be of the same order.
Nondimensionalizing the system (\ref{vpfpe}) -
(\ref{norma}) in these units, we can rewrite it
as
\begin{eqnarray}
\nabla_{v}\cdot [( v + \nabla_x \phi)\,\rho +
\nabla_v \rho] = \epsilon\,\left(\frac{\partial
\rho}{\partial t} + v\cdot \nabla_x
\rho\right) \,, \label{dvpfpe}\\ 
\Delta_{x}\phi =
- \int \rho\, dv, \label{dpoisson2}\\ 
\int \rho\,
dx\, dv = 1, \label{dnorma}
\end{eqnarray}
to be solved together with appropriate initial
conditions and decay boundary conditions at
infinity.

We are interested in finding a simpler reduced
equation for $P(x,t)$ in the high-field limit as
$\epsilon\to 0+$. This problem has been recently
tackled by Nieto, Poupaud and Soler, \cite{NPS},
who proved rigorously that $P$ obeys the following
hyperbolic system (in dimension one):
\begin{eqnarray}
P_{t} = {\partial\over\partial x}\, ( \phi_x
P),\label{1}\\
\phi_{xx} = - P, \label{2}\\
\int P\, dx =1, \label{3}
\end{eqnarray}
to leading order as $\epsilon\to 0+$ (subscripts
mean partial derivation with respect to the
indicated variable). By using the decay of $P$ at
infinity, they derived the following equation
for the electric field, $E= -\phi_x$:
\begin{eqnarray}
E_t + E\, E_x = 0.  \label{4}
\end{eqnarray}
As is well-known, this equation may develop shock
waves in finite time. Nieto {\it et al} proved that
there exists a unique entropy solution of the VPFP system
in the high-field limit, and that the shock
velocity corresponds to interpreting (\ref{4})
as the conservation law $E_t + {1\over 2}
(E^2)_x = 0$. The uniqueness result suggests that
this conservation law could be regularized by
adding a small viscosity term, $\epsilon\, 
E_{xx}$.  Then analysis of the resulting
Burgers equation in the limit as $\epsilon\to 0+$
yields the unique entropy solution. Their method
amounted to closing the hierarchy of equations
for the moments of $\rho$, and leaves open the
question of how to add higher-order terms to the
system (\ref{1}) - (\ref{3}), or (which is
related) how to regularize (\ref{4}).

In this paper, we shall consider these problems.
To this end, we shall compare three classical
approaches in kinetic theory, the Hilbert
expansion, closure of moment equations and the
Chapman-Enskog method (CEM) as applied to the
one-dimensional VPFP system. These approaches
yield the same result for the parabolic scaling
\begin{eqnarray}
\nabla_v \cdot ( v \,\rho + \nabla_v \rho) =
\epsilon\,
\left(v\cdot \nabla_x \rho - \nabla_x
\phi\cdot \nabla_v \rho\right) +
\epsilon^2\,\frac{\partial
\rho}{\partial t}\,,   \label{5}
\end{eqnarray}
which is interesting and usually employed in
kinetic theory \cite{chapman,cercignani}. This
scaling corresponds to lower values of the
electric field, so that the diffusive and
frictional terms in the VPFP equation are in fact
much larger than the corresponding term in 
(\ref{vpfpe}). For the present model, the parabolic
scaling was also studied rigorously  by Poupaud and
Soler \cite{PS}, who derived from (\ref{5}) a
drift-diffusion for $P=\int \rho\, dv$ in the limit
as $\epsilon\to 0+$. In the high-field limit
corresponding to the scaling (\ref{dvpfpe}), we
shall see that only the CEM yields satisfactory
results {\em systematically}. In fact, the Hilbert
expansion yields the following additional system
for the order
$\epsilon$ correction to $P$, $P^{(1)}$:
\begin{eqnarray}
P^{(1)}_{t} - {\partial\over\partial x}\, (
\phi^{(0)}_x P^{(1)} + \phi^{(1)}_x P) =
P_{xx},\label{6}\\
\phi^{(1)}_{xx} = - P^{(1)}, \label{7}\\
\int P^{(1)}\, dx =0. \label{8}
\end{eqnarray}
At face value, this equation breaks down in the
shock regions. However, it is compatible with the
result of the CEM,
\begin{eqnarray}
P_{t} - {\partial\over\partial x}\, (
\phi_x P) = \epsilon\, P_{xx},\label{9}\\
\phi_{xx} = - P, \label{10}\\
\int P\, dx = 1, \label{11}
\end{eqnarray}
provided we substitute $P$ by $P+\epsilon
P^{(1)}$ and equate like powers of $\epsilon$ in
both sides of these equations. This situation is
exactly that found when using the method of
multiple scales to describe certain codimension
two bifurcations \cite{BPS,bon00}. The key point
is that the CEM yields reduced equations which
may contain terms of different order in
$\epsilon$, unlike the other methods. Appropriate
closure of the equations for the moments of
$\rho$ may yield (\ref{9}) - (\ref{11}), but it
is not a systematic method.

The rest of the paper is as follows. The Hilbert
expansion and the CEM are applied to the
one-dimensional VPFP system in Sections
\ref{sec:2} and \ref{sec:3}, respectively. The
latter section also contains the result of
applying the Chapman-Enskog procedure to the
three-dimensional problem. Section \ref{sec:4}
shows how to obtain the same results by closing
the hierarchy of equations for the moments.
Finally Section \ref{sec:5} contains our
conclusions.

\section{The Hilbert expansion}
\label{sec:2}
It consists of inserting the power series
\begin{eqnarray}
\rho = \rho^{(0)} + \epsilon \rho^{(1)} +
\epsilon^2 \rho^{(2)} + O(\epsilon^3), \label{12}
\end{eqnarray}
in the VPFP system. This expansion is really akin
to the method of multiple scales with slow time
scale $t$ \cite{kev96}. We assume that the leading
order contribution to the solution has already
relaxed in the fast time scale. Equating like
powers of
$\epsilon$ in both sides of the one-dimensional
VPFP system, we obtain
\begin{eqnarray}
\frac{\partial}{\partial v}\left\{
\frac{\partial\rho^{(0)}}{\partial v} +
[v + \phi^{(0)}_{x}]\, \rho^{(0)}\right\} = 0,
\label{13}\\
\frac{\partial^{2}\phi^{(0)}}{\partial x^{2}} = -
\int_{-\infty}^{\infty} \rho^{(0)}\, dv,
\label{14}\\
\int_{-\infty}^{\infty} \rho^{(0)} dx dv =1,
\label{15}
\end{eqnarray}

\begin{eqnarray}
{\cal L}\rho^{(1)}\equiv \frac{\partial}{\partial
v}\left\{ \rho^{(1)}_{v} + [v + \phi^{(0)}_{x}]\,
\rho^{(1)} + \phi^{(1)}_{x}]\, \rho^{(0)} \right\}
\nonumber\\
= \rho^{(0)}_t + v \rho^{(0)}_x,
\label{16}\\
\phi^{(1)}_{xx} = - \int_{-\infty}^{\infty}
\rho^{(1)}\, dv, \label{17}\\
\int_{-\infty}^{\infty} \rho^{(1)} dx dv = 0,
\label{18}
\end{eqnarray}

\begin{eqnarray}
{\cal L}\rho^{(2)} = \rho^{(1)}_t + v
\rho^{(1)}_x - \phi^{(1)}_x \rho^{(1)}_v,
\label{19}\\
\phi^{(2)}_{xx} = - \int_{-\infty}^{\infty}
\rho^{(2)}\, dv, \label{20}\\
\int_{-\infty}^{\infty} \rho^{(2)} dx dv = 0,
\label{21}
\end{eqnarray}
and so on.

The solution of (\ref{13}) - (\ref{15}) is
\begin{eqnarray}
\rho^{(0)} = {e^{-{V^{2}\over
2}}\over\sqrt{2\pi}}\, P(x,t),\label{22}
\end{eqnarray}
for a function $P$ to be determined and such that
\begin{eqnarray}
V = v + \phi^{(0)}_x, \label{23}\\
\phi^{(0)}_{xx} = - P ,\label{24}\\
\int_{-\infty}^{\infty} P(x,t)\, dx = 1.
\label{25}
\end{eqnarray}
Notice that the Fokker-Planck collision term
behaves more nicely than the general Boltzmann
term. In the high-field limit, the latter may give
rise to runaway solutions for many forms of the
collision frequency \cite{frosali}.

Since $\int {\cal L}\rho^{(n)} dv = 0$, (\ref{16})
yields the following solvability condition 
$${\partial\over\partial t}\int \rho^{(0)} dv +
{\partial\over\partial x}\int v\rho^{(0)} dv= 0.$$
necessary for $\rho^{(1)}$ to be bounded even as
$v\to \pm\infty$. This provides
\begin{eqnarray}
P_{t} = {\partial\over\partial x}\, (\phi^{(0)}_x
P),\label{26}\\
\phi^{(0)}_{xx} = - P, \label{27}
\end{eqnarray}
together with (\ref{3}), which coincides with
Nieto {\it et al}'s result, \cite{NPS}.

Suppose we now want to correct the leading order
result by going one step further, to solving
(\ref{16}) - (\ref{18}). The result is
\begin{eqnarray}
\rho^{(1)} = {e^{-{V^{2}\over 2}}\over\sqrt{2
\pi}}\, \left( P^{(1)} - [P_x + \phi^{(1)}_x
P]\, V \right.\nonumber\\
\left. - {V^{2}-1\over 2}\, P^{2} \right)\,,
\label{28}\\
\phi^{(1)}_{xx} = - P^{(1)} , \label{29}\\
\int P^{(1)} dx = 0, \label{30}
\end{eqnarray}
where $P^{(1)}(x,t)$ is yet to be determined. By
using the solvability condition for (\ref{19}), we
obtain
\begin{eqnarray}
P^{(1)}_{t} - {\partial\over\partial x}\,
(\phi^{(0)}_x P^{(1)} +\phi^{(1)}_x P)
= P_{xx},\label{31}
\end{eqnarray}
to be solved together with (\ref{29}) and
(\ref{30}). Notice that the left hand side of
(\ref{31}) is a linearization of (\ref{26}) about
$P$. This equation does not make sense at those
points where $P$ is discontinuous, but we can
easily see that (\ref{26}) and (\ref{31}) are
obtained by taking ${\cal P} \sim P + \epsilon
P^{(1)}$ in the following equation:
\begin{eqnarray}
{\cal P}_t &-& {\partial\over\partial x}\,
\left( \phi_x\, {\cal P} + \epsilon {\cal
P}_{x}\right) = O(\epsilon^2)\,, \label{32}
\end{eqnarray}
to be solved together with the Poisson equation
$\phi_{xx} = - {\cal P}$ and $\int {\cal P} dx =
1$. This situation is exactly that found when
using the method of multiple scales to find the
normal form describing certain codimension two
bifurcations \cite{BPS,bon00}. The Hilbert
expansion is akin to the method of multiple
scales in that it yields reduced equations all
whose terms are of the same order. Thus if we
insist in calculating higher-order equations, we
obtain coupled systems of equations such as those
written above. To find directly a reduced
equation including terms of different order in
$\epsilon$, we should use the CEM. Appropriate
closure of the equations for the moments of
$\rho$ may yield (\ref{32}), but not in a
systematic manner.

\section{Chapman-Enskog method}
\label{sec:3}
In one dimension, the dimensionless VPFP system is
\begin{eqnarray}
\frac{\partial}{\partial v}\left\{ \frac{\partial\rho}{\partial v}
+ [v + \phi_{x}]\, \rho\right\} = \epsilon\,\left(\frac{\partial
\rho}{\partial t} + v\frac{\partial\rho}{\partial x}\right) \,,
\label{-dfpe}\\ \frac{\partial^{2}\phi}{\partial x^{2}} = -
\int_{-\infty}^{\infty} \rho\, dv, \label{dpoisson1}
\end{eqnarray}
to be solved together with
\begin{equation}
\int_{-\infty}^{+\infty}\int_{-\infty}^{+\infty} \rho(x,v,t)\,
dx\, dv = 1, \label{dnorma1}
\end{equation}
and appropriate initial and decay conditions as
$v$ and $x$ tend to $\pm \infty$.

\subsection{Chapman-Enskog method}
Setting $\epsilon=0$ in (\ref{-dfpe}), we find a simple equation
to be solved together with (\ref{dpoisson1}) and (\ref{dnorma1}).
Its solution is a displaced Maxwellian:
\begin{eqnarray}
\rho = {e^{-{V^{2}\over 2}}\over\sqrt{2\pi}}\,
P(x,t),\label{ce1}\\ V = v + \phi^{(0)}_x, \label{ce2}\\
\phi^{(0)}_{xx} = - P ,\label{ce3}\\ \int_{-\infty}^{\infty}
P(x,t)\, dx = 1. \label{ce4}
\end{eqnarray}
Notice that $P(x,t)$ is an arbitrary
function of $x$ and $t$ except for (\ref{ce4}).
Furthermore, (\ref{ce1}) correspond to a
particular choice of initial conditions. The
Chapman-Enskog ansatz consists of assuming that
$\rho$ has the following asymptotic expansion
\begin{eqnarray}
\rho = {e^{-{V^{2}\over 2}}\over\sqrt{2\pi}}\,
P(x,t;\epsilon) + \sum_{n=1}^{\infty}
\epsilon^n \, \rho^{(n)}(x,v;P).
\label{ce5}
\end{eqnarray}
Furthermore, we impose that the amplitude $P$
obeys an equation:
\begin{eqnarray}
{\partial P\over\partial t} = \sum_{n=0}^{\infty}
\epsilon^n \, F^{(n)}(P),\label{ce6}
\end{eqnarray}
where $F^{(n)}$ are functionals of $P$ to be
determined as the procedure goes on. This
equation for $P$ is not explicitly written in the
usual presentations of CEM
\cite{chapman,cercignani}. Instead, the form of
this equation is guessed by writing equations for
the moments of $\rho$ and using gradient
expansions. We find this latter procedure more
confusing.

Insertion of (\ref{ce5}) and (\ref{ce6}) into the
equations and auxiliary conditions yields a
hierarchy of linear equations for the
$\rho^{(n)}$. Notice that the latter depend on
time only through their dependence on
$P$. The functionals $F^{(n)}(P)$ are determined
so that each equation (and set of auxiliary
conditions) for $\rho^{(n)}$ has a solution which
is bounded for all values of $v$, even as
$v\to\pm\infty$. Once a sufficient number of
$F^{(n)}$ is determined, (\ref{ce6}) {\em is the
sought amplitude equation}. Please notice that,
unlike results from the method of multiple
scales, terms in (\ref{ce6}) may be of different
order.

Let us illustrate how the procedure works by finding $F^{(0)}$ and
$F^{(1)}$. Insertion of (\ref{ce5}) and (\ref{ce6}) in
(\ref{-dfpe}), (\ref{dpoisson1}) and (\ref{dnorma}) yields the
following hierarchy of linear equations:
\begin{eqnarray}
{\cal L}\rho^{(1)} &=& {e^{-{V^{2}\over 2}}\over\sqrt{2\pi}}\,
\left[v P_x - v V \phi^{(0)}_{xx} P + F^{(0)} 
\right.\nonumber\\
 &-& \left.
VP\phi^{(0)}_{xt}\big|_{P_{t}= F^{(0)}} \right]\,,
\label{ce7}\\  
{\cal L}\rho^{(2)} &=& {e^{-{V^{2}\over 2}}\over
\sqrt{2\pi}}\, \left[F^{(1)} - VP
\phi^{(0)}_{xt}\big|_{P_{t}= F^{(1)}}
\right] + \rho^{(1)}_t \nonumber\\
&+& v \rho^{(1)}_x -
\phi^{(1)}_x\, \rho^{(1)}_v  \, , \label{ce8}
\end{eqnarray}
and so on. Before equating like powers of
$\epsilon$, we have substituted $P_t =
\sum_{n=0}^\infty F^{(n)}$ when performing time
differentiations such as $\phi^{(0)}_{xt}$. This
yields the obvious terms in the hierarchy of
equations, which results in rather cluttered
formulas, as we ascend in the hierarchy of
equations. These equations are to be supplemented
by the normalization conditions (\ref{ce4}), 
\begin{eqnarray}
\int_{-\infty}^{\infty} \rho^{(n)} dv = 0, \ n 
\geq 1, \label{ce9}
\end{eqnarray}
and the linear equations and definitions:
\begin{eqnarray}
\phi^{(n)}_{xx} = - \int_{-\infty}^{\infty} \rho^{(n)} dv\,,
\label{ce10}\\ {\cal L}\rho^{(n)} = {\partial\over\partial v}\,
\left[V\rho^{(n)} + \rho^{(n)}_{v} + P \, \phi^{(n)}_{x}\,
{e^{-{V^{2}\over 2}}\over\sqrt{2\pi}}\right] \,, \label{ce11}
\end{eqnarray}
for $n=1,2,\ldots$. $V$ is again given by
(\ref{23}).

Let us now consider (\ref{ce7}). Since the $v$
integral of its left hand side is zero, this
equation has a solution only if the $v$ integral
of its right hand side is zero. The corresponding
integrals are simplified by using the symmetry of
the Maxwellian and shifting integration variables
from $v$ to $V$. The condition that the integral
of the right side vanish yields
\begin{eqnarray}
F^{(0)} =\left\{\phi^{(0)}_x\, P \right\}_x .
\label{ce12}
\end{eqnarray}
Notice that we needed $F^{(0)}$ in the right side
of (\ref{ce7}) for this equation to have an
appropriate solution.

We now calculate $\phi^{(0)}_{xt}$ in order to
simplify the right side of (\ref{ce7}). As we
explained above, at this order we should substitute
$P_t = F^{(0)} =\left\{\phi^{(0)}_x\, P \right\}_x$
when needed. Hence
\begin{eqnarray}
\phi^{(0)}_{xt} = - K_x * P_t = - K_x *
\left\{\phi^{(0)}_x\, P \right\}_x \nonumber\\
= - K_{xx} *
\left\{\phi^{(0)}_x\, P
\right\} = - \phi^{(0)}_x\, P,\label{calculo}
\end{eqnarray}
where $K$ is the Green's function of the
one-dimensional Laplacian and $*$ means convolution
product. In the previous calculation, we have
integrated by parts and used the decay properties
of $P$ and the symmetry of $K$. Inserting now
(\ref{ce12}) and (\ref{calculo}) in Eq.\
(\ref{ce7}), we realize that the right hand side
thereof is the partial derivative of
$(V\phi^{(0)}_{xx}P-P_x)\, e^{-V^{2}/2}/
\sqrt{2\pi}$ with respect to $v$. This immediately
yields
\begin{eqnarray}
\rho^{(1)} &=& - {e^{-{V^{2}\over
2}}\over\sqrt{2\pi}}\, \left( {V^{2}-1\over 2}\,
P^2  + V P_x\right)\, , \label{ce13}
\end{eqnarray}
which satisfies (\ref{ce9}) and yields $\phi^{(1)}
=0$. In (\ref{ce13}), we have omitted adding a
term like that proportional to $P^{(1)}$ in the
right hand side of (\ref{28}) (satisfying $\int
P^{(1)}\, dx =0$) because all such terms are
already included in the amplitude
$P(x,t;\epsilon)$. 

To find $F^{(1)}$, we insert (\ref{ce13}) in
(\ref{ce8}) and use the solvability condition for
this equation. Simplifications arise from the
identities
$$\int_{-\infty}^{\infty} \rho^{(1)}_t dv =
{\partial\over\partial t} \int_{-\infty}^{\infty}
\rho^{(1)} dv = 0,
$$
and
$$\int_{-\infty}^{\infty} v \rho^{(1)}_x
dv =  {\partial\over\partial x}
\int_{-\infty}^{\infty} v
\rho^{(1)}\, dv.
$$
The result is
\begin{eqnarray}
F^{(1)} = P_{xx} \,.  \label{ce14}
\end{eqnarray}

We can now insert (\ref{ce12}) and (\ref{ce14})
into (\ref{ce6}) to obtain the sought Smoluchowski
equation for $P$:
\begin{eqnarray}
P_t &-& {\partial\over\partial x}\,
\left( \phi_x\, P + \epsilon P_{x}\right)
= 0\,, \label{ce15}
\end{eqnarray}
to be solved together with (\ref{ce3}).
Substituting (\ref{ce3}) into (\ref{ce15}), and
integrating with respect to $x$, we find the
following Burgers equation for $E= - \phi_x$:
\begin{eqnarray}
E_t + E\, E_x = \epsilon\, E_{xx}\,.  \label{ce16}
\end{eqnarray}
Provided that the initial electric field be
uniformly bounded, this Burgers equation has
solutions which converge to the distributional
solutions of the Hopf equation (\ref{4}). Thus the 
CEM is in fact a vanishing-viscosity method which
yields a unique solution approaching the unique
entropy solution of (\ref{4}) as $\epsilon\to 0+$.
In Ref.\ \cite{NPS}, the authors proved convergence
towards the unique entropy solution of the Hopf
equation by means of a different method which used
that the limiting electric field is a monotone
decreasing function.

\subsection{Three-dimensional VPFP system}
The CEM consists of inserting the following
ansatz in (\ref{dvpfpe})
\begin{eqnarray}
\rho = {e^{-{V^{2}\over 2}}\over
(2\pi)^{{3\over 2}}}\, P(x,t;\epsilon) +
\sum_{n=1}^{\infty}\epsilon^n \,
\rho^{(n)}(x,v;P). \label{ce17}\\
{\partial P\over\partial t} = \sum_{n=0}^{\infty}
\epsilon^n \, F^{(n)}(P),\label{ce18}\\
V = v + \nabla_x \phi. \label{ce19}
\end{eqnarray}
Then $\phi$ obeys the Poisson equation
\begin{eqnarray}
\Delta_x \phi = - P,\label{ce20}
\end{eqnarray}
and the normalization conditions are
\begin{eqnarray}
\int P(x,t)\, dx = 1, \label{ce21}\\
\int\rho^{(n)}(x,v;P)\, dx dv = 0.\label{ce22}
\end{eqnarray}
Equations (\ref{ce7}) and (\ref{ce8}) are obviously generalized to
the three-dimensional case. The solvability condition for the
first equation yields $$F^{(0)} = \nabla_x\cdot (\nabla_x \phi\,
P).$$ 
Substituting $F^{(0)}$ in (\ref{ce7}) we find
\begin{eqnarray}
{\cal L}\rho^{(1)} = \nabla_v \cdot \left\{
{e^{-{V^{2}\over 2}}\over (2\pi)^{{3\over 2}}}\,
\left[P\, v. \nabla_x \nabla_x \phi^{(0)} \right.
\right.\nonumber\\
\left.\left. - \nabla_x P + P\nabla_x \phi^{(0)}_t 
\right] \right\}.  \nonumber
\end{eqnarray}
The solution is then
\begin{eqnarray}
\rho^{(1)} &=& {e^{-{V^{2}\over 2}}\over
(2\pi)^{{3\over 2}}}\, \left\{ {P\over 2}\, (VV -
I): \nabla_x \nabla_x\phi^{(0)} \right.\nonumber\\
&+& \left. V\cdot \, \left[P\, \nabla_x
\left(\phi^{(0)}_t - {|\nabla_x
\phi^{(0)} |^{2}\over 2} \right) - \nabla_x
P\right] \right\} .
\label{ce23}
\end{eqnarray}
Notice that in the one-dimensional case, the term
$\nabla_x \left(\phi^{(0)}_t - {|\nabla_x
\phi^{(0)} |^{2}\over 2} \right) $ vanishes. We
could have added a term which solves the
homogeneous equation ${\cal L} u=0$ in the right
hand side of (\ref{ce23}). However, as we explained
before, the effect of this term is already included
in $P(x,t;\epsilon)$. The solvability condition for
the $\rho^{(2)}$ equation yields
\begin{eqnarray}
F^{(1)} = \nabla_x\cdot\left\{P\, \nabla_x \left(
{|\nabla_x \phi^{(0)} |^{2}\over 2} - \phi^{(0)}_t
\right) + \nabla_x P \right\}\,. \label{ce24}
\end{eqnarray}
The reduced Smoluchowski equation for $P$ is
therefore
\begin{eqnarray}
P_t = \nabla_x \cdot\left\{P\nabla_x \phi +
\epsilon\, P\, \nabla_x
\left( {|\nabla_x \phi|^{2}\over 2} - \phi_t
\right) \right.\nonumber\\
\left. + \epsilon\,\nabla_x P
\right\}\,, \label{ce25}
\end{eqnarray}
up to terms of order $\epsilon^2$. Since $P=
\nabla_x\cdot E$, $E=-\nabla_x \phi$, we may
rewrite this equation as
\begin{eqnarray}
E_t + E (\nabla_x \cdot E) = \epsilon\, \left\{
(\nabla_x\cdot E)\, [(E\cdot\nabla_x) E +
E_t ] \right.\nonumber\\
\left. + \nabla_x (\nabla_x\cdot E)\right\} +
{\cal J}\,, \label{ce26}
\end{eqnarray}
where $\nabla_x\cdot {\cal J} = 0$, so that ${\cal
J}$ is a solenoidal field. In the one-dimensional
case, ${\cal J} = 0$, $E_{t} = - E E_x +
O(\epsilon)$, and (\ref{ce26}) becomes the Burgers
equation (\ref{ce16}).

\section{Closure of equations for the moments}
\label{sec:4}
Equations (\ref{ce15}) or (\ref{ce16}) can also
be obtained by appropriate closure of the
hierarchy of equations for the moments of $\rho$,
as we will show now. The equations for the first
two moments of $\rho$,
$$P = \int \rho\, dv, \quad J = \int v\rho\, dv,
$$
can be obtained directly from the VPFP system:
\begin{eqnarray}
P_t &+& J_x = 0,\label{c1}\\
-J &-& (\phi_x P)_x = \epsilon J_t +
\epsilon\, {\partial\over\partial x}\,
\int v^2 \rho\, dv.\label{c2}
\end{eqnarray}
If we take the $x$-derivative of the second
equation and eliminate $J$ by using the first
one, we obtain
\begin{eqnarray}
P_t - (\phi_x P)_x = - \epsilon P_{tt} +
\epsilon\, {\partial^{2}\over\partial x^{2}}\,
\int v^2 \rho\, dv.\label{c3}
\end{eqnarray}
Let us now close the hierarchy of equations by
inserting the zeroth order ansatz (\ref{ce1})
\begin{eqnarray}
\int v^2 \rho\, dv \sim \int v^2 {e^{-{V^{2}\over
2}}\over\sqrt{2\pi}}\, P\, dv = (1+\phi^2_x)\, P
\label{c4}
\end{eqnarray}
in (\ref{c3}). The result is the nonlinear wave
equation \cite{jbk}:
\begin{eqnarray}
P_t - (\phi_x P)_x = \epsilon\,
\left\{ {\partial^{2}\over\partial x^{2}}\,
[(1+\phi^2_x)\, P] - P_{tt}\right\}\,,
\label{c5}
\end{eqnarray}
to be solved together with (\ref{ce3}) and
(\ref{ce4}). For the electric field, $E=-\phi_x$,
$P=E_x$, and (\ref{c5}) yields
\begin{eqnarray}
E_t + E\, E_x + \epsilon\, E_{tt} -
\epsilon\, {\partial\over\partial x}\,
[(1+E^2)\, E_x] = 0, \label{c6}
\end{eqnarray}
after integrating once with respect to $x$ and
using the decay conditions to cancel the
resulting constant. We can now show that this
equation is compatible with our previously
derived Burgers equation. Let us iterate
(\ref{c6}) to get
\begin{eqnarray}
E_{tt} \sim - (E\, E_x)_t = - E_t E_x - E E_{xt}
\nonumber\\
\sim E E_x^2 + (E\, E_x)_x\, E = (E^2 E_x)_x\,.
\label{c7}
\end{eqnarray}
Inserting this result in (\ref{c6}), we recover
the Burgers equation (\ref{ce16}). In the limit
as $\epsilon\to 0+$, both the Burgers equation
and Equation (\ref{c6}) regularize shock waves and
yield the same shock speed, $(E_+ + E_-)/2$, for a
shock connecting the field values $E_-$ and
$E_+$ \cite{whitham}. However the inner structure
of the shock is different in both cases and it
would be interesting to study Equation (\ref{c6})
for its own sake.

\section{Conclusions}
\label{sec:5}
We have derived a reduced equation for the charge
density (or the electric field) of the VPFP
system in the high-field limit. The simplicity of
the Fokker-Planck collision term allows us to
explicitly find a zeroth order solution which is
a displaced Maxwellian. This result is the basis
of two classical singular perturbation procedures
used to obtain the reduced equation: the Hilbert
expansion and the Chapman-Enskog method. We find
that the second method is ideally suited to
provide a reduced equation whose terms may be of
different order. In the present case, we obtain
a form of the Burgers equation. On the other
hand, closure of the equations for the moments
of the probability density yields a nonlinear
wave equation, which is compatible with the
previous result, but may be of independent
interest. Another interesting open problem is to
use the CEM in the rigorous analysis of the
high-field limit of the VPFP system. So far,
rigorous analyses have used the less systematic
method of closing moment equations. The trouble
with using CEM seems to be that it is more
``nonlinear'' (albeit systematic) than the other
methods. \\

\noindent {\bf Acknowledgements}
LLB thanks Prof.\ J.\ B.\ Keller for showing him the
derivation of the nonlinear wave equation in
Section IV. We thank Mr.\ J.\ Nieto and Profs.\ C.\
Cercignani and F.\ Poupaud for helpful comments.
This work was supported by the Spanish DGES through
grants PB98-0142-C04-01, PB98-1281 and by the
European Union TMR contract ERB FMBX-CT97-0157.


\end{multicols}
\end{document}